\documentclass{an}

\usepackage{times}
\usepackage{epsfig}

\def \etal   {\hbox{et~al.\/}}

\def\lesssim{\mathrel{\hbox{\rlap{\hbox{\lower4pt\hbox{$\sim$}}}\hbox{$<$}}}}
\def\gtrsim{\mathrel{\hbox{\rlap{\hbox{\lower4pt\hbox{$\sim$}}}\hbox{$>$}}}}

\begin{document}

\title{Recombination Lines and Free-Free Continua Formed in Asymptotic Ionized
Winds: Analytic solution for the radiative transfer }

\author{Richard Ignace\thanks{Email:  ignace@etsu.edu}}

\titlerunning{The FREC model}
\authorrunning{R. Ignace}

\institute{Department of Physics \& Astronomy,
East Tennessee State University,
Box 70652,
Johnson City, TN 37614,
USA}

\received{}
\publonline{}

\abstract{
In dense hot star winds, the infrared and radio continua
are dominated by free-free opacity and recombination
emission line spectra.  In the case of a spherically
symmetric outflow that is isothermal and expanding
at constant radial speed, the radiative transfer for
the continuum emission from a dense wind is analytic.
Even the emission profile shape for a recombination
line can be derived.  Key to these derivations is that
the opacity scales with only the square of the density.
These results are well-known.  Here an extension of the
derivation is developed that also allows for line blends
and the inclusion of an additional power-law dependence
beyond just the density dependence.  The additional
power-law is promoted as a representation of a radius
dependent clumping factor.  It is shown that differences
in the line widths and equivalent widths of the emission
lines depend on the steepness of the clumping power-law.
Assuming relative level populations in LTE in the upper
levels of He\,{\sc ii}, an illustrative application of the
model to {\em Spitzer}/IRS spectral data of the carbon-rich
star WR~90 is given.
}

\maketitle

\section{Introduction}

The winds of massive stars are exceedingly dense, especially in evolved
stars such as OB supergiants, Luminous Blue Variables (LBVs), and
Wolf-Rayet (WR) stars (e.g., Cassinelli 1979; Lamers \& Cassinelli 1999;
Kudritzki \& Puls 2000).  The winds are dense and highly ionized, to the
point where free-free opacity can become optically thick in the wind not
only at radio wavelengths but also in the infrared (IR).  Consequently,
the IR~band offers opportunities to study the wind density distribution
through continuum and line emission.  Wright \& Barlow (1975) showed
that at radio wavelengths the specific flux of emission for a spherical,
constant expansion, isothermal, and optically thick wind will produce a
continuum power-law having a slope of $f_\nu \propto \nu^{0.6}$, where
$f_\nu$ is the specific flux (e.g., in Janskys).  Knowing the distance
to the star, the specific luminosity is derivable and can be related to
the wind mass-loss rate $\dot{M}$.  Observations of appropriate sources
at radio frequencies have been important for deriving $\dot{M}$ values
(e.g., Abbott \etal\ 1980; Leitherer, Chapman, \& Koribalski 1995).

Understanding mass loss during the various evolutionary stages of massive
stars is of critical importance, as for example in understanding observed
ratios of blue and red supergiant stars or the relative numbers of WR
subtypes, the nitrogen, carbon, and oxygen rich WR stars (Maeder \&
Meynet 2000; Meynet \& Maeder 2003).  Mass-loss rates are also important
for interpreting the afterglows of gamma-ray bursts, of which some are
associated with supernova explosions (Woosley \& Bloom 2006).

Abbott, Bieging, \& Churchwell (1981) showed that radio emission
used to infer mass-loss rates can also depend on ``clumping'' if stellar
winds are not laminar or ``smooth''.  The reason is that the free-free
opacity scales with the square of the wind density.  As a result,
more or less clumping biases the mass-loss rate determinations if
clumping corrections are not taken into account, with ramifications for
our understanding of massive star evolution.  And not only will clumping
affect mass-loss rates, it can even have feedback for the line-driving
efficiency of the winds in some cases (Brown \etal\ 2004; Oskinova,
Hamann, \& Feldmeier 2007).

It has become abundantly clear that stellar winds are indeed clumped, and
that the clumping is not negligible in relation to interpreting $\dot{M}$
values.  Already, researchers had good reason to expect clumping to be
important, since the hot star winds are understood to be line-driven
(Castor, Abbott, \& Klein 1975; Friend \& Abbott 1986; Pauldrach, Puls,
\& Kudritzki 1986), a mechanism that is subject to instability leading to
the formation of wind structure (Lucy \& Solomon 1970; Owocki, Castor,
\& Rybicki 1988; Feldmeier, Puls, \& Pauldrach 1997).  Evidence for
clumping in WR star winds and now even O star winds has been mounting for
many years.  In the case of WR stars, a convincing argument was made by
Hillier (1991) on the basis of the emission of recombination line cores
in relation to the strength of the electron scattering wings of the lines.
The former is a density square process and subject to influence by
clumping, but the latter is linear in density.  Smooth wind models were
found inadequate in matching the line profile shapes, but the inclusion
of clumping allowed for superior matches between models and data.

Information about the clumping in the less dense winds of O stars
has come only more recently (Bouret \etal\ 2003; Evans \etal\ 2004;
Bouret, Lanz, \& Hillier 2005).  Indeed, quite startling results have
been reported by Fullerton, Massa, \& Prinja (2006) who suggest that on
the basis of FUV doublet lines of P\,{\sc v} that early O star mass-loss
rates might need downward revisions by factors of 10 to perhaps 100.
This result along with the general recognition of the prevalence of
clumping in massive star winds led to an entire meeting devoted to the
topic (Hamann, Feldmeier, \& Oskinova 2007).

Many researchers have shifted attention from the question of whether
the wind clumping exists to its origin and evolution (i.e., how clumping
initiates, how it evolves through the wind, and how its nature changes
with stellar evolutionary phase).  A better understanding of clumping
requires fresh considerations of what diagnostics may be used to constrain
clump properties, such as size and mass distribution and dynamics.
For example, Nugis, Crowther, \& Willis (1998) considered how a gradient in the
clumping factor with radius in the wind modifies continuum slopes for
the wind emission.  The main point of their study is that a constant
clumping factor affects the amount of emission, but it does {\it not}
change the observed power-law slope of the continuum.  Only a clumping
factor that is radius dependent will influence the slope.

Using the results of hydrodynamic simulations, Runacres \& Owocki
(2002, 2005) have considered how a structured flow evolves in radius for
1-dimensional wind models.  Necessarily, such models predict clumping in
the form of spherical shells.  This has long known not to be an accurate
description of hot star winds, for example owing to the relatively
low variability observed in the X-ray emissions from massive stars
(Cassinelli \& Swank 1983; Bergh\"{o}fer 1997).  Indeed, there have been
attempts to model the low X-ray variability in terms of statistically
spherical but 3-dimensionally structured wind flow (Oskinova \etal\ 2001).
Although impressive progress has been made in extending the 1-dimensional
numerical simulations to an axisymmetric geometry (Dessart \& Owocki
2002, 2003, 2005), the problem of fully 3-dimensional and time-dependent
radiative hydrodynamic models remains a challenge.

With so much emphasis being given to measuring wind clumping factors
and understanding structured wind flow, useful insight and tools can be
gained from considering some simplified cases.  As was noted,
Wright \& Barlow (1975) were able to derive a power-law continuum slope
of index $f_\nu \propto \nu^{0.6}$ at radio wavelengths for a spherical,
isothermal, and inverse square law density.  Still
assuming a smooth wind flow, Cassinelli \& Hartmann (1977)
extended the approach to allow for power-law density and temperature
distributions.  They derived analytic expressions for the continuum
slope.  And indeed, IR~spectra of WR~winds are known to be steeper than
the canonical $\nu^{0.6}$ rule (e.g., Morris et al.\ 1993), and so
approaches like that of Cassinelli \& Hartmann are relevant.

Emission line profile shapes have also been considered.  Hillier, Jones,
\& Hyland (1983) derived an analytic solution for the radiative transfer
of the line and continuum emission for density-squared opacities.
For the continuum part, the standard Wright \& Barlow (1975) solution
is recovered in terms of the dependence on the wind parameters and the
power-law scaling in frequency.  The inclusion of continuous opacity and
line optical depth in the Sobolev approximation yields a line profile
shape that is centrally peaked and gently rounded in appearance (see
Ignace, Quigley, \& Cassinelli 2003).

Here the Hillier \etal\ (1983) solution is extended in two significant
ways.  First, their derivation prescribes a $r^{-2}$ dependence for the
wind density, but an additional power-law factor in radius can also be
included in the opacity.  Such a power law may represent the effects
of a radius-dependent clumping factor.  Second, it turns out that the
radiative transfer can allow for an {\it arbitrary} number of line blends,
which is important for IR spectra dominated by emission lines.  The result of
the new formulation is an analytic solution to a fairly complex scenario
of line blends and power-law opacity but in the restrictive limit of
spherical symmetry and constant expansion.  Although direct applications
may be limited, the solution provides complex cases against which sophisticated
radiative transfer routines can be benchmarked.

The following section describes the approach to the analytic solution
that extends the results of Hillier \etal\ (1983), following the notation
of Ignace \etal\ (2003).  Section~\ref{sec:app} provides an illustrative
application to high resolution spectra of the carbon-rich WC star, WR~90.
A summary of results is given in section~\ref{sec:summ}.

\section{Radiative Transfer Solution}

\subsection{The Physical Ingredients}

The scenario being addressed assumes spherical symmetry for a wind
expanding at constant radial speed.  For a fixed mass-loss rate, this
means that the density is an inverse square law, hence 

\begin{equation}
\rho = \frac{\dot{M}}{4\pi\,v_\infty\,r^2}.
	\label{eq:density}
\end{equation}

\noindent Only continuum and line processes that scale with the square
of the density will be considered.  This is a reasonable approximation
for dense, ionized winds of hot stars where free-free continuum opacity
and recombination line opacity are important. The wind will be assumed
isothermal.  For the line transfer, the Sobolev approximation will
be adopted.

Keys to the continuum radiative transfer are the expressions for the
continuum and line optical depths.  For free-free opacity, the optical
depth from the observer to a point in the wind is

\begin{equation}
\tau_{\nu,{\rm ff}} = R_*\,K_{\rm ff}(T,\nu)\, \int_z^\infty
	n_{\rm i}\,n_{\rm e}\, dz',
	\label{eq:ff}
\end{equation}

\noindent where $R_*$ is the radius of the star, $n_{\rm i}$ is the
ion number density, $n_{\rm e}$ is the electron number density, and $T$
is the isothermal temperature.  Coordinates normalized to the stellar
radius are used, with $(p,\phi, z)$ for cylindrical coordinates
and $(x,\theta,\phi)$ for spherical coordinates.  The factor $K_{\rm ff}$
is given by (Cox 2000):

\begin{equation} 
K_{\rm ff} = 3.692\times 10^8 \,\left( 1-e^{-h\nu/kT}\right)
	Z_{\rm i}^2 g_\nu T^{-1/2}\nu^{-3},
\end{equation}

\noindent where $g_\nu$ is the Gaunt factor and $Z_{\rm i}$ is the
root mean square ion charge.  For the Gaunt factor at IR wavelength, a
power-law in frequency is adopted as given by Carciofi \& Bjorkman (2006),

\begin{equation}
g_\nu = g_0\,(\nu/\nu_0)^{-u},	\label{gaunt}
\end{equation}

\noindent with $u \approx 0.23$ at a frequency $\nu_0$ corresponding
to 1~$\mu$m.  Note that the number density quantities relate to the
mass density in terms of mean molecular weights per ion $\mu_{\rm i}$
and per free electron $\mu_{\rm e}$, respectively.  Hence $\mu_{\rm i}
m_H n_{\rm i} = \mu_{\rm e}m_H n_{\rm e} = \rho$.

The next step is to consider the line transfer, for which the
Sobolev approximation
is used.  Although the wind is in constant radial expansion, a
line-of-sight gradient in the projected line-of-sight velocity occurs
owing to the divergent expansion of the spherical wind.  Since highly
supersonic wind speeds are of interest, the Sobolev approximation
remains valid except for the extreme line wings where spatially the
velocity gradient vanishes along the sightline from the observer to the
star center.

For constant radial expansion, the observed Doppler shift in velocity
units along the $z$-axis is

\begin{equation}
v_{\rm z} = - v_\infty \, \cos \theta = -v_\infty \, \mu,
\end{equation}

\noindent where $\theta$ is the spherical polar angle from the $z$-axis as
previously noted, and $\mu = \cos \theta$.  The main value of the Sobolev
approximation is the identification of emission at a fixed velocity shift
in an observed line profile with a spatial locus of points in the flow (or
``isovelocity zones'').  In this case a fixed velocity shift transforms
to a conical surface in the spherical wind.

Sobolev theory does more than simply identify isovelocity zones.  It also
offers a solution to the line transfer (e.g., Castor 1970; Rybicki \&
Hummer 1978; Hummer \& Rybicki 1983, 1985).  It is known that Sobolev
theory for modeling line profiles under the assumption of a smooth hot
star wind is inadequate for producing realistic fits to some spectral
features, such as the ``black troughs'' of ultraviolet P~Cygni lines
(Prinja, Barlow, \& Howarth 1990).  The problem is that not only are hot
star winds clumpy in density, but the radial profile of the wind velocity
is non-monotonic (Lucy \& Solomon 1970; Lucy 1982).  As a practical matter
of line profile synthesis, the radiative transfer has been addressed by
some in the form of a modified Sobolev approach, such as the SEI method of
Lamers, Cerruti-Sola, \& Perinotto (1987) or the co-moving frame methods
of Hillier \& Miller (1998) or Gr\"{a}fener, Koesterke, \& Hamann (2002).
The non-monotonic behavior is approximated as an average smooth wind with
substantial ``turbulent'' broadening (e.g., Groenewegen \& Lamers 1989).

Since the Sobolev approximation is related to velocity
gradients, it may be inappropriate for wind clumps if gradients
are absent throughout such structures.  However, the situation may not
be so grave for the Sobolev approximation.  First, a good model for the
3-dimensional structure (density and velocity) is currently lacking,
and so the absence of velocity gradients remains a question.  Second,
in the scenario of this paper, the flow is treated as being in constant
expansion, and the velocity gradient reduces to a geometrical one
(see below) owing to spherical divergence, and clumps can certainly be
expected to participate in the overall bulk radial expansion.  Finally,
the Sobolev approximation should remain valid for clumps that are small
compared to the Sobolev length (i.e., $L \sim v_{\rm br}/(dv_{\rm z}/dr)$
for $v_{\rm br}$ the dominant local broadening process) and that are
optically thin (in contrast to treatments of thick clumps as in Oskinova
\etal\ 2007). Although limitations for the line transfer in the Sobolev
approach are noted, Sobolev is employed to the restricted problem at hand
to explore the influence of clumping on emission lines
and continuum formation at long wavelengths.

The key parameter for describing the line transfer is the Sobolev
optical depth $\tau_S$ for a line with opacity at point $(p, \phi, z)$;
the optical depth is given by

\begin{equation}
\tau_S = \frac{\kappa_L\,\rho\,\lambda_0\, R_*}{|dv_{\rm z}/dz|},
\end{equation}

\noindent where $\kappa_L$ is the frequency integrated line opacity 
with dimensions of
area per mass times frequency, $\lambda_0$ is the central wavelength of
the line transition, and the denominator is the line-of-sight velocity
gradient evaluated at the isovelocity zone for fixed impact parameter $p$.
The parameter $R_*$ appears because $z$ is a normalized coordinate.
In spherical symmetry the velocity gradient becomes

\begin{equation}
\frac{dv_{\rm z}}{dz} = \frac{dv}{dx}\,\cos^2\theta + \frac{v}{x}\,\sin^2
	\theta,
	\label{eq:vgradient}
\end{equation}

\noindent where $x=r/R_*$.  Equation~(\ref{eq:vgradient}) reduces to
$(v_\infty/x)\,(1-\mu^2)$ for a wind in constant expansion.

\begin{figure}
\centering{\epsfig{figure=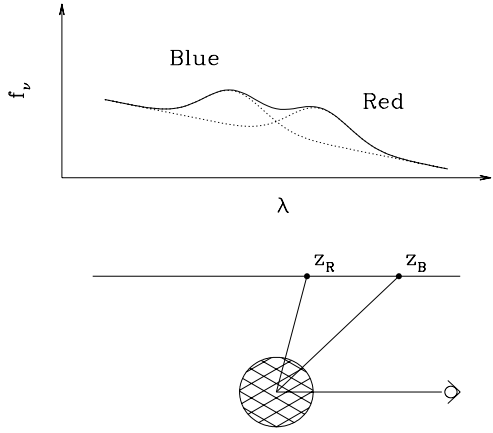, height=6.8cm}}
\caption{\small Lower:  A sightline that passes through the wind.
The two points are intersections of the sightline with two different
isovelocity cones shown here in cross-section.  The geometry is for
a line blend, such that at fixed $\lambda$, the transition with
line center redward (``R'') of $\lambda$ is the rearward point,
and the one that is blueward (``B'') is nearer the observer.
Above is a cartoon illustration of a line blend that might be
emerge from the wind, with solid for the total and dotted for
the individual line contributions.
\label{fig2}}
\end{figure}

\subsection{The Solution to the Radiative Transfer}
\label{subsec:solution}

\subsubsection{Solution for General Line Source Functions}

Before embarking on the steps of the derivation, more parameters must be
introduced to describe the properties of the radiation.  The continuum
source function is represented by $S_C$, which for free-free processes
is Planckian, hence $S_C = B_\nu (T)$.  The line source function will
be $S_L$.

The derivation largely follows the notation of Ignace \etal\ (2003)
who elaborated on the steps originally described by Hillier \etal\
(1983).  For $S_L=S_C=B_\nu(T)$, the
total emergent intensity of continuum and line radiation is given by
(eq.~[A1] of Ignace \etal):

\begin{equation}
I_\nu = B_\nu \, \left[ 1 - e^{-(\tau_{\rm max}+\tau_S)}\right],
	\label{eq:Idumb}
\end{equation}

\noindent where $\tau_S$ is evaluated at the point $(p,z)$, and $\tau_{\rm
max}$ is the maximum free-free optical depth along a sightline. 
The assumption of an isothermal wind is implicit in equation~(\ref{eq:Idumb})
so that $B_\nu$ may be factored out of the integral to obtain a solution
to the transfer equation.

As in Ignace \etal, it is convenient to express continuum and line
optical depths in terms of scale parameters $\tau_{\rm C}$ and $\tau_L$
that encompass the various physical properties of the wind and opacity
that do not vary with spatial location (such as $T$, $\nu$, etc).
The Sobolev and optical depth can be expressed as:

\begin{equation}
\tau_S = \tau_L\,x^{-3}\,(1-\mu^2) = \tau_L\,\sin\theta\,p^{-3}.
\end{equation}

\noindent which uses the fact that $p=x\sin \theta$.  The continuum
free-free optical depth from the observer to any point $z$ in the wind
along a fixed sightline of impact parameter $p$ derives from an integral
along the path as given by

\begin{equation}
\tau_{\rm ff} = \tau_{\rm C} \,\int_z^\infty\,\frac{dz}{x^4\,w^2(x)}.
\end{equation}

\noindent Eliminating $x$ again in favor of $p$ and $\theta$, the integral
for the continuum optical depth has an analytic solution given by

\begin{equation}
\tau_{\rm ff} = \frac{\tau_{\rm C}}{2p^3}\,(\theta-\sin\theta\,\cos\theta) 
\end{equation}

\noindent where $\tan \theta = p/z$.  The parameter $\tau_{\rm C}$ is
a scaling parameter for the free-free optical depth that incorporates
the various wind and atomic constants involved along with the frequency
dependence; in relation to equation~(\ref{eq:ff}), it is given by:

\begin{equation}
\tau_{\rm C} = R_\ast\,K_{\rm ff}(T,\nu)\,n_{\rm i,0}\,n_{\rm e,0},
	\label{eq:tc1}
\end{equation}

\noindent where $n_{\rm i,0}$ and $n_{\rm e,0}$ are the ion and electron
number densities at the wind base $R_\ast$.
The maximum optical depth $\tau_{\rm max}$ along a sightline that does
not intercept the stellar photosphere is

\begin{equation}
\tau_{\rm max}= \frac{\pi\tau_{\rm C}}{2p^3}.
\end{equation}

When $\tau_{\rm ff} \gg 1$, an IR/radio pseudo-photosphere forms that is
much larger than the hydrostatic stellar photosphere.  In this case there
is a standard integral that allows for the integration of the emergent
intensity with impact parameter to derive the total flux of line and
continuum radiation (Hillier \etal\ 1983).  The new
considerations include line blends, an additional power-law dependence
to the opacity, and line source functions that need not be Planckian.
Although a closed form for the emergent intensity can be given for
this new generalization, there is unfortunately no analytic solution
for the total flux, except in the special case of LTE for the line
source functions.  (That case will be considered in subsequent sections.)

In deriving the emergent intensity, the first of the new considerations
is to allow for a line blend.  Suppose there is a wavelength where
two separate lines contribute to the opacity.  The two lines shall
be referenced in terms of respective line center wavelengths as the
``red'' line (centered at $\lambda_R$) and the ``blue'' line (centered at
$\lambda_B$) relative to the wavelength $\lambda$ under consideration.
As illustrated in Fig.~\ref{fig2}, the velocity shift for the red line
is necessarily at an isovelocity zone that lies rearward of that for the
blue line.  Figure~\ref{fig2} shows a sightline as it might cross the
two isovelocity zones (and no more!) for the two respective lines at
wavelength $\lambda$.  The following equation describes contributions
from alternating emission and attenuations by the respective line and
continuum opacities in the different segments of the figure as working
from left to right:

\begin{eqnarray}
I_\nu & = & \;\;\; B_\nu\,\left( e^{-\tau_R} - e^{-\tau_{\rm max}} \right) \,
	e^{-\tau_{\rm S,R}}\,e^{-\tau_{\rm S,B}} \nonumber\\ 
 & &	+ S_R\,\left( 1 - e^{-\tau_{\rm S,R}} \right) \, e^{-\tau_R}\,e^{-\tau_{\rm S,B}} \nonumber \\
 & &	+ B_\nu\,\left( e^{-\tau_B}-e^{-\tau_R} \right) \, e^{-\tau_{\rm S,B}}
	\nonumber \\ 
 &  &  + S_B\,\left( 1-e^{-\tau_{\rm S,B}}\right) \,e^{-\tau_B}  \nonumber \\
 &  &  + B_\nu\,\left( 1-e^{-\tau_B}\right) , 
	\label{eq:ILONG}
\end{eqnarray}

\noindent There is an identifiable pattern that can be extended to allow
for blends of an arbitrary number of lines.  It is convenient to split
the emergent intensity into two parts:  the continuum contribution $I_C$
arising from terms with the continuum source function and $I_L$ involving
contributions from all of the line source functions.  

The emergent intensity from the free-free continuum emission is given by

\begin{eqnarray}
I_C & = & B_\nu \, \left[ 1 - \sum_{j=1}^{N+1}\,e^{-\tau_j}\,
	\exp\left(-\sum_{i=0}^{i<j}\,\tau_{{\rm S,}i}\right) \right. \nonumber\\ 
    &  & \left. +\sum_{j=1}^{N}\,e^{-\tau_j}\,
	\exp\left(-\sum_{i=1}^{i=j}\,\tau_{{\rm S,}i}\right)\right],
	\label{eq:IC}
\end{eqnarray}

\noindent where $N$ is the total number of lines contributing to
the blend at a fixed frequency in the spectrum.  In this formulation
$\tau_i$ is the free-free continuum optical depth from the observer
to the $i^{\rm th}$ resonance point along a given ray through the wind.
In this notation $\tau_{N+1} = \tau_{\rm max}$.  Then $\tau_{{\rm S,}i}$
is the Sobolev optical depth at the $i^{\rm th}$ resonance point, with
$\tau_{\rm S,0} \equiv 0$.

The contribution of the total line intensity is given by

\begin{equation}
I_L = \sum_{i=1}^{N}  S_i \left( 1-e^{-\tau_{{\rm S,}i}} \right)
	e^{-\tau_i}\exp\left(-\sum_{j=0}^{j-1}
	\tau_{{\rm S,}j} \right).
	\label{eq:IL}
\end{equation}

\noindent Note that at this point, the only requirement on the solution
for equations~(\ref{eq:IC}) and (\ref{eq:IL}) is that the wind be
isothermal; velocity structure, ionization gradients, or variable clumping
factors are allowed in the set of continuum optical depths $\tau_j$.
The set of line optical depths $\tau_{{\rm S,}i}$ are also quite general
since they are locally evaluated only at crossings of the sightline with
isovelocity zones.  Indeed, it is not even necessary at this point to
specify the velocity law or the geometry of the isovelocity zones; that
information is only required when evaluating the total flux of emission.
Given that $B_\nu$ is constant, the above relations result only from a
consideration of sequencing.

The total emergent intensity along the ray is simply $I_\nu = I_{\rm C}
+ I_{\rm L}$.  In order to obtain the continuum spectral distribution
and line profile shapes, additional simplifying assumptions of constant
expansion and $S_{{\rm L,}i}= B_\nu$ are introduced next as a particular
realization of possible wind models.

\subsubsection{The Solution for LTE Source Functions}

It is well-known that the level populations of atomic species in stellar
winds is highly NLTE (Auer \& Mihalas 1972).  However, the application
being considered here is for IR/radio spectra involving H\,{\sc i} or
He\,{\sc ii} recombination lines and free-free continua.  The continuum
that is formed at large radius and thus roughly coincident with the line
formation is LTE.  The lines themselves are for transitions involving
high $n$-level values whose populations should be governed largely by
recombination and photoionization processes.  These lines are formed
at sufficiently large radius that in the case of the WR winds, the
EUV radiation from the star is (a) highly diluted and (b) strongly
absorbed by the dense wind.  In this case the photoionization from the
spatially local IR~continuum that is Planckian (although not isotropic)
will contribute to the populations of the upper levels.  Indeed,
Hillier \etal\ (1983) found that the relative upper level populations
for some high $n$-levels of He\,{\sc ii} in EZ~CMa were consistent with LTE.  

Here LTE is invoked as a convenience to allow for an analytic
solution to the line and continuum fluxes and quick exploration of the
model parameters.  There is a tremendous simplication for the emergent
continuum and line intensities, because all but two terms cancel exactly.
After simplifying expression~(\ref{eq:ILONG}) using $S_{\rm R}=S_{\rm
B}=B_\nu$, the emergent intensity reduces to just

\begin{eqnarray}
I_\nu & = & B_\nu \, \left[ 1 - e^{-(\tau_{\rm max}+\tau_{\rm S,R}+\tau_{\rm S,B})}
	\right] ,		\label{eq:Itwo} \\
 & = & B_\nu \, \left[ 1 - e^{-\tau_{\rm T}}
        \right] .               \label{eq:Itot}
\end{eqnarray}

\noindent where $\tau_{\rm T}$ is a sum of all three optical depths.
As before, this can be extended for an arbitrary number of line blends,
now with $\tau_{\rm T} = \tau_{\rm max} + \sum_{i}^{N} \,
\tau_{{\rm S,} i}$.  Each Sobolev optical depth must be evaluated for
its respective isovelocity zone at a fixed frequency.  This amounts
to different velocity shifts for lines of different wavelength centers
$\{\lambda_i\}$.  These shifts can be written in normalized form as
$w_i = v_{{\rm z},i}/v_\infty = - \mu_i$.  It is the total flux
of emission that is required to simulate observables, and the intensity
is now of a form that can be solved analytically because both the line
and continuum optical depths have the same radial dependence.

Another important extension to previous treatments is that an {\em
additional} power-law dependence of the opacity can be included.  The
winds are known to be clumped, and so this new power-law dependence can
be used to represent a radius dependence of the clumping.  The clumping
factor $D_C$ is introduced as follows:

\begin{equation}
D_C = \frac{\langle\rho^2\rangle}{\langle \rho \rangle^2} \propto x^{-m}.
\end{equation}
                                                                                
\noindent As a result, the Sobolev optical depth becomes

\begin{eqnarray}
\tau_S & = & \tau_L \, x^{-(3+m)}\,(1-\mu^2)^{-1} \\
 & = & \tau_L \, p^{-(3+m)}\,\left(\sin \theta\right)^{1+m}, \label{eq:tS}
\end{eqnarray}
                                                                                
\noindent where $\tau_L$ is a scale factor for the optical depth.
Equation~(\ref{eq:tS}) in terms of $p$ and $\theta$ will
prove useful in evaluating the total flux.  The free-free optical
depth becomes

\begin{eqnarray}
\tau_{\rm max} & = & \tau_{\rm C} \, \int_{-\infty}^{\infty}\, x^{-(4+m)}\,dz \\
 & = & \tau_{\rm C}\,p^{-(3+m)}\,\int_0^\pi\,(\sin \theta)^{2+m} \, d\theta\label{tauCmid} \\ 
 & \equiv &  \tau_{\rm C}\,p^{-(3+m)}\times\gamma(m). \label{eq:tC}
\end{eqnarray}

\noindent where $\tau_{\rm C}$ from equation~(\ref{eq:tc1}) is now
expressed as

\begin{equation}
\tau_{\rm C} = \frac{K_{\rm ff}(T,\nu) \, \dot{M}^2}{16\pi^2\,R_*^3\,v_\infty^2\,
	\mu_{\rm e}\,\mu_{\rm i}\,m_H^2}
	\label{eq:tc2}
\end{equation}

\noindent and $\gamma$ is a constant that depends solely on the power-law
index $m$ via the integral factor of equation~(\ref{tauCmid}).  In the
case that $m=0$, $\gamma = \pi/2$.  Formally, the latter expression only
applies for rays that do not intersect the star, hence only for $p> 1$.

As discussed by Wright \& Barlow (1975), as long as the continuum
optical depth is quite large, a pseudo-photosphere is formed that is
signficantly larger in cross-section than the star itself.  In this
case ignoring rays with $p<1$ represents a small error in the flux.
Under these circumstances, the solution for the total flux of emission
in the continuum and lines is analytic.  Even with line blends and the
additional power-law factor, the form of the integral is similar to that
of Ignace \etal\ (2003).  Here the steps for deriving $f_\nu$
are reviewed.

The flux is given by an integral of the intesity over all rays: 

\begin{equation}
f_\nu = 2\pi\,\frac{R_\ast^2}{d^2}\,\int_\nu\,I_\nu(p)\,p\,dp,
\end{equation}

\noindent where the integral accounts for only those spatial points in
the wind that contribute to emission for some frequency in the
spectrum $\nu$.  Unlike the single line case considered in Ignace
\etal, this frequency point may have contributions from multiple
line opacity sources.  Using equation~(\ref{eq:Itot}) with constant
$B_\nu$, the flux becomes

\begin{equation}
f_\nu = 2\pi B_\nu\,\frac{R_\ast^2}{d^2}\,\int_0^\infty\,
	\left(1-e^{-\tau_{\rm T}(p)}\right)\,p\,dp,
\end{equation}
                                                                                
\noindent where $d$ is the distance to the star,
and from equations~(\ref{eq:tS}) and (\ref{eq:tC}) the total
optical depth is now

\begin{eqnarray}
\tau_{\rm T} (p) & = & p^{-(3+m)}\,\times \nonumber \\
 & & \left[\tau_{\rm C}\,\gamma(m)
	+ \sum_{i}^{N}\,\tau_{{\rm L},i}\,(\sin\theta_i)^{1+m} \right].
\end{eqnarray}

\noindent Again, the use of the form of the total optical depth above
in the preceding integral relation for the flux assumes that the flux
contribution for rays with $p<1$ is relatively small.  This is a good
approximation at long wavelengths because the effective photosphere can
indeed be much larger than $R_\ast$.  The advantage of this treatment
is that the integral is analytic.

\begin{figure*}
\centering{\epsfig{figure=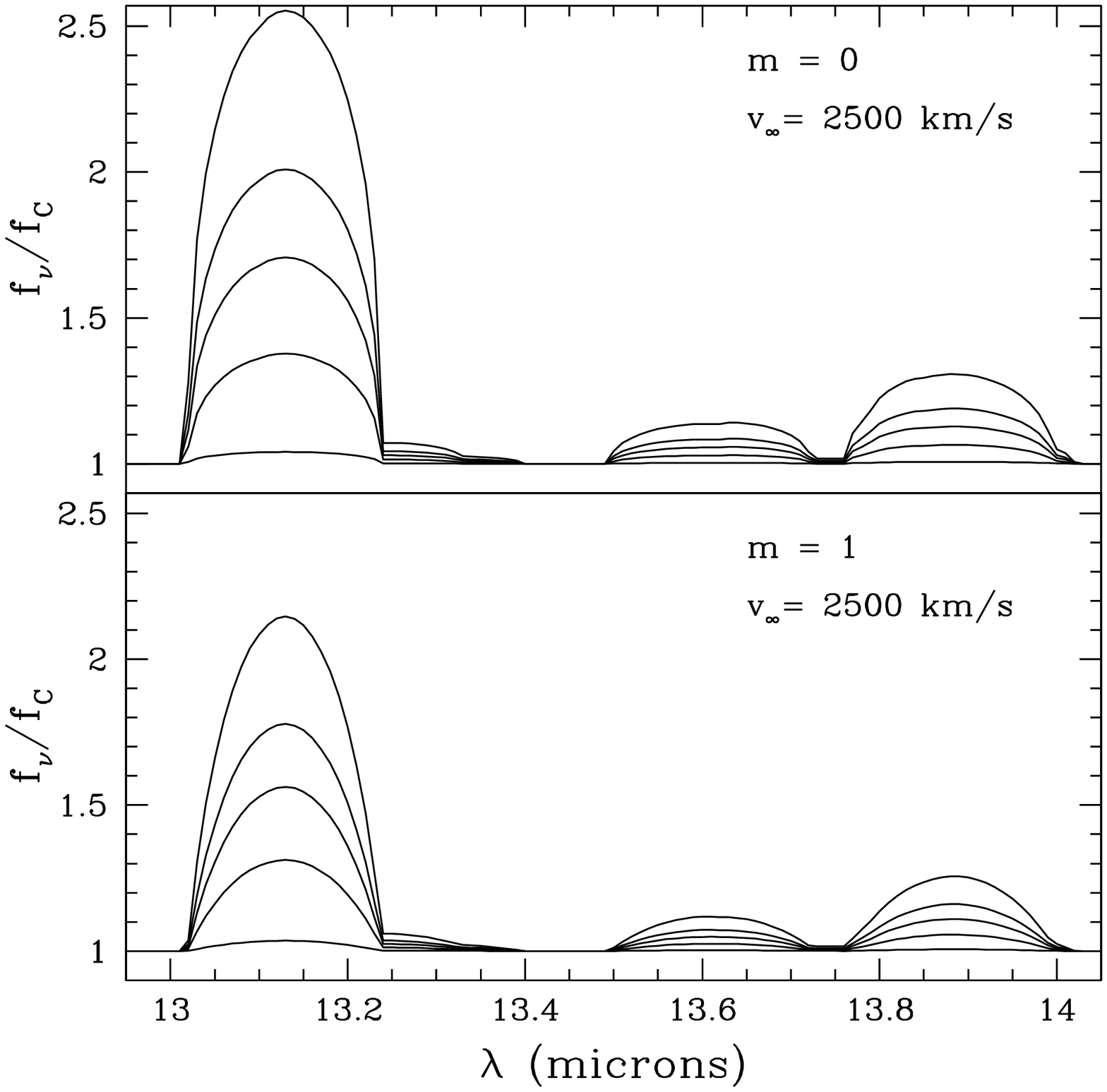, height=8.0cm}\epsfig{figure=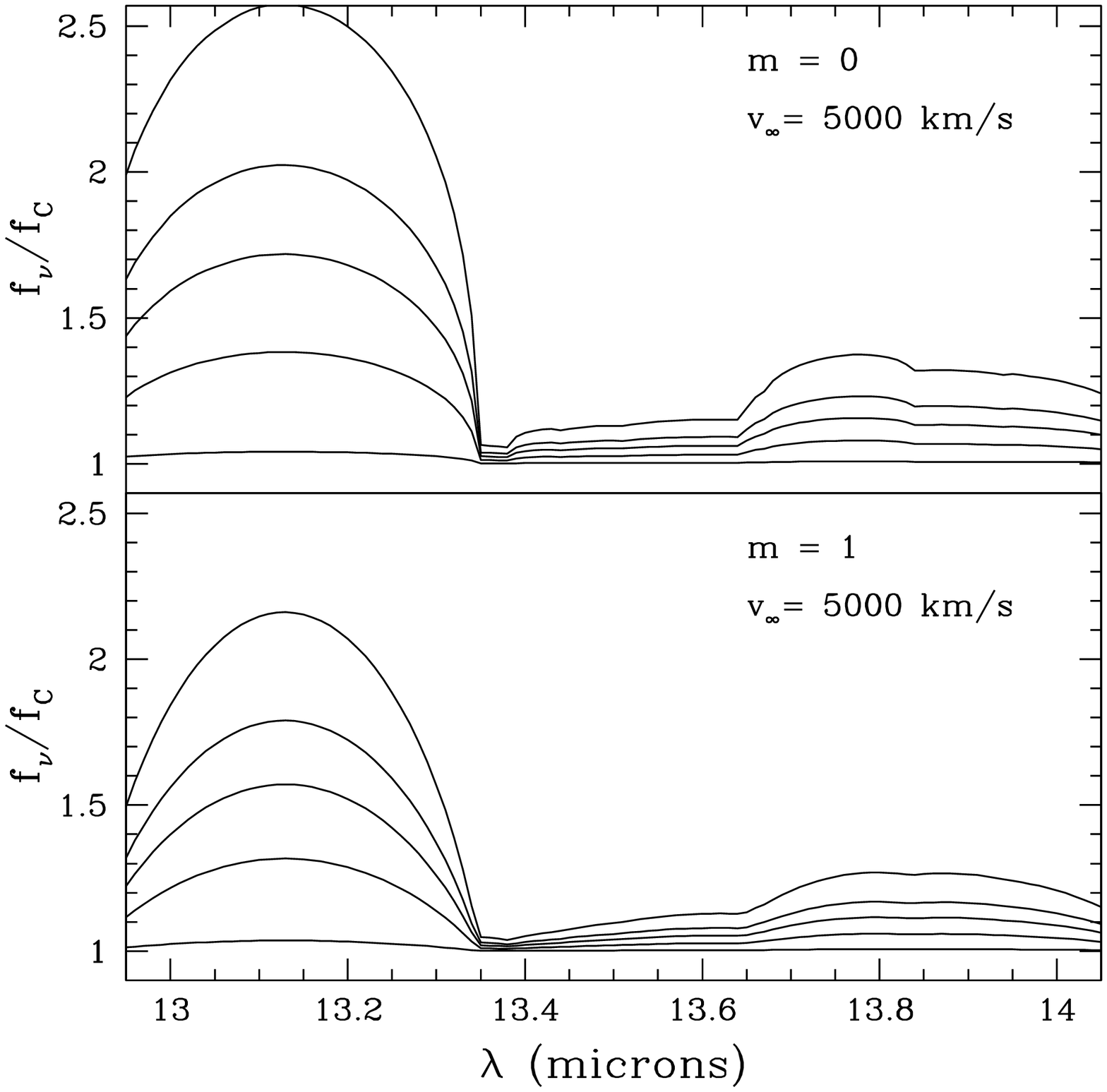, height=8.0cm}}
\caption{\small
Illustration of the analytic solution for a continuum dominated by
free-free emission and He\,{\sc ii} recombination lines, such as occurs
for WR~stars.  Spectra are shown as normalized to the continuum $f_C$.
The two plots left and right are for two different wind terminal speeds
as indicated.  The upper and lower panels are for two different power
law indices $m$:  upper is for constant clumping factor and lower is for
clumping that decreases inversely with radius.  Note that the analytic
results account fully for line blending.  The different curves within
each panel are for different line strengths. \label{fig3}}
\end{figure*}

To arrive at the solution, it is useful to introduce a change of variable
with $U^{(3+m)} = \tau_{\rm C}\,p^{-(3+m)}$ to convert the integral
expression to the following form:

\begin{eqnarray}
f_\nu & \approx & 2\pi B_\nu\,\frac{R_\ast^2}{d^2}\,\int_0^\infty\,
	\tau_{\rm C}^{2/(3+m)}\,U^{-3}\,dU \,\times \nonumber \\
 & & \Big\{1-\exp\big[U^{3+m}\,(\, \gamma \big. \Big. \nonumber \\
 & & \Big.\big.  - \tau_{\rm C}^{-1}\,\sum_{i=0}^{N} \tau_{{\rm L},i} \sin^{m+1} 
	\theta_i ) \big]\Big\},
	\label{eq:fluxint}
\end{eqnarray}

\noindent where the sum refers to any of the lines contributing to
the opacity at the frequency of interest.  For $N=0$ there are no
contributing line opacities, $\tau_{{\rm L},0} =0$, and expressions
reduces to an expression for the continuum emission alone.

The integral in that form of (\ref{eq:fluxint}) has a solution given in
Gradshteyn \& Ryzhik (2000) found in section 3.478, \#2, with the final
result for the observed flux $f_\nu$ being

\begin{eqnarray} 
f_\nu & = & 2\pi B_\nu\,\frac{R_\ast^2}{d^2}\,
\tau_{\rm C}^{2/(3+m)}\, \left[ \frac{3+m}{2}\;
        \Gamma\left(\frac{1+m}{3+m}\right)\right] \nonumber \\
 & &  \times\,\Bigg\{\, \gamma(m) \nonumber \Bigg.\\
 & &  \Bigg. + \sum_{i=0}^{N}\,\frac{\tau_{{\rm L,}i}}{\tau_{\rm C}}\,
        \left(1-w_i^2\right)^{(m+1)/2} \Bigg\}^{2/(3+m)},
        \label{eq:FREC}
\end{eqnarray}

\noindent where $\Gamma$ is the Gamma-function, and $\sin
\theta = (1-w_{\rm z}^2)^{1/2}$. For reference, the solution
equation~(\ref{eq:FREC}) shall be referred to as the ``Free-free +
Recombination Emission line in Constant velocity'' model, or the
``FREC'' model.

In regions of the spectrum that are absent of lines, the FREC model
provides the continuum flux.  In the Rayleigh-Jeans limit, the continuum
reduces exactly to a power-law form.  Assuming $h\nu \ll kT$ and using
the Gaunt factor of equation~(\ref{gaunt}), the continuum distribution
between lines becomes a function of just one free parameter $m$ (i.e.,
for the value of $u$ known):

\begin{equation} 
f_\nu \propto \nu^{(2+2m)/(3+m)}\,g_\nu^{2/(3+m)} \propto \nu^{2(m+1-u)/(3+m)}.
        \label{eq:fluxscale}
\end{equation}

\noindent Note that with $u$ appropriate for the radio band, and $m=0$, the
$0.6$ power-law slope of Wright \& Barlow (1975) is recovered.   

Returning to the lines, why does the line shape vary with $m$-value?
Although both line and continuum opacities scale with the square of the
density, the line optical depth has an additional dependence with angular
escape through a velocity gradient.  As a result, the line shape is not
invariant with respect to the continuum formation as $m$ is varied.

Examples of model spectra produced by the FREC solution are shown
in Fig.~\ref{fig3}.  The models are continuum-normalized to focus on
the line shape effects and the inclusion of blending.  The left and
right plots are for different wind speeds of 2500 and 5000 km~s$^{-1}$.
All panels show a segment of the 13--14 micron spectrum.  All emission
lines are of He\,{\sc ii} recombination; the strong line at 13.12 microns
is He\,{\sc ii} 10$\alpha$.  The upper panel is for $m=0$, corresponding
to a wind that has a constant clumping factor.  The lower panel is for
$m=1$, indicating a clumping factor that decreases with radius.

To clarify better the line-shape dependence on the index $m$, it is useful
to consider limiting cases of a single strong or weak line: a strong
line occurs for $\tau_L \gg \tau_{\rm C}$ and a weak one for $\tau_L
\ll \tau_{\rm C}$.  A single strong line yields the profile shape given by

\begin{equation} 
f_\nu \propto \left[ 1-w^2 \right]^{(1+m)/(3+m)}.
\end{equation}

\noindent 
In the case of
a weak line, the flux is of the form $f_\nu \propto 1 + \epsilon$, where
$\epsilon$ has the functional dependence of

\begin{equation}
\epsilon \propto \left[ 1-w^2 \right]^{(1+m)/2}.
\end{equation}
                                                                                
The profile shapes for thin and thick lines are not the same.  Thick lines
tend to be more ``rounded'' or ``bubbled'' as compared to thin ones.
The half width at half maximum (HWHM) of a thick line is greater than
for a thin one.  In the case of a thick line, it is

\begin{equation}
{\rm HWHM}_{\rm thick} = v_\infty \,\sqrt{1 - \left(\frac{1}{2}\right)^{(3+m)/(1+m)} } ,
\end{equation}

\noindent whereas for a thin line, it is

\begin{equation}
{\rm HWHM}_{\rm thin} = v_\infty \,\sqrt{1 - \left(\frac{1}{2}\right)^{2/(1+m)} } .
\end{equation}

\begin{figure}
\centering{\epsfig{figure=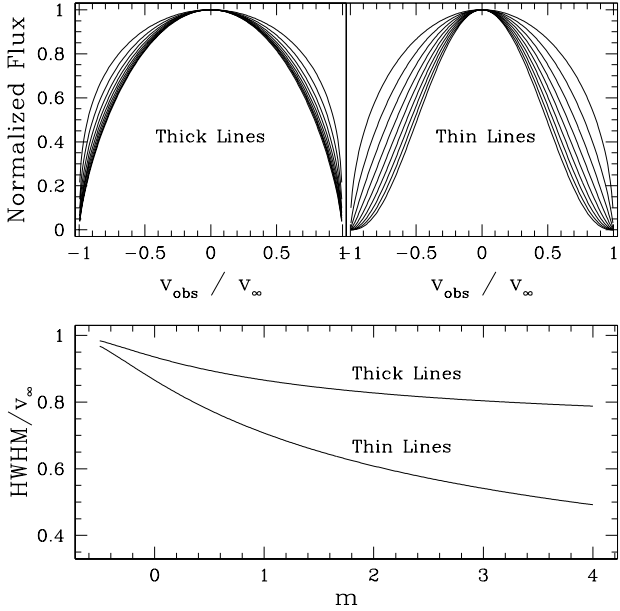, height=8.6cm}}
\caption{\small Comparisons between thick and thin emission lines
from the FREC model.  The two top panels display line profiles
for thick (left) and thin (right) lines.  These are continuum
subtracted and then normalized to their peak values and
plotted against normalized velocity shift.  Each
curve is for a different $m$ value, from $m=-0.5$ to 4 in
intervals of $0.5$.  At bottom is a plot of the HWHM of thick
and thin lines against $m$.  The line widths are normalized
to the wind terminal speed.
\label{fig4}}
\end{figure}

\noindent For $m=0$, HWHM$_{\rm thick} = 0.94v_\infty$ and HWHM$_{\rm
thin}= 0.87v_\infty$.  The difference in half-width from the full
terminal speed is small, but it becomes greater for increasing values
of $m$.  With $m=1$, the half-width values become $0.87 v_\infty$ versus
$0.71v_\infty$.  For fast winds of WR~stars at around 2000 km~s$^{-1}$,
the difference amounts to over 300 km~s$^{-1}$, which is potentially
measurable.  Figure~\ref{fig4} compares, as a function of $m$ value, thick
versus thin line profile shapes (top two panels) with a corresponding plot
(bottom panel) of the HWHM values relative to the wind terminal speed.

\begin{figure}
\centering{\epsfig{figure=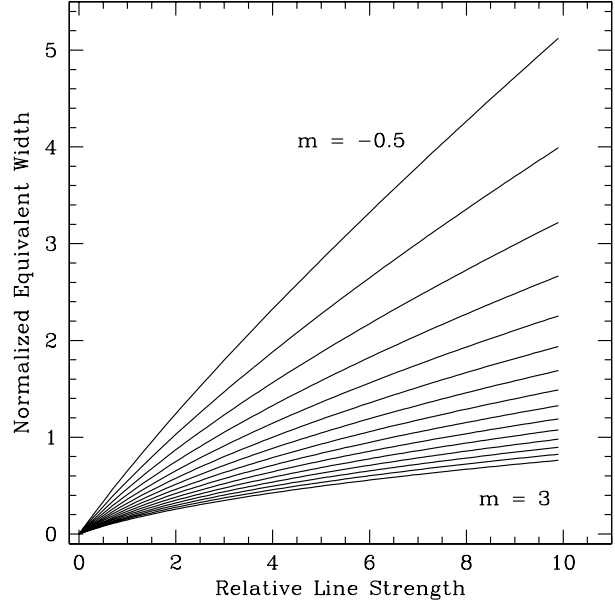, height=8.6cm}}
\caption{\small Normalized line equivalent widths (EW) are
plotted against the relative line strength (see text).
Each curve is for a different power-law index $m$, ranging
from $m=-0.5$ up to $m=3$ in increments of 0.25.  The EW
is seen to grow in line strength as expected, but the rapidity of
growth is much less for clumping factors that are more steeply
decreasing with radius.
\label{fig4b}}
\end{figure}

It is also useful to consider trends in the line equivalent width ($EW$).
The line equivalent width for the FREC model with a single
line is given by:

\begin{eqnarray}
EW & = & \frac{\lambda_0\,v_\infty}{c} \times \\ \nonumber
 & & \int_{-1}^{+1} \left[ 1 + \frac{\tau_L}
	{\gamma(m)\,\tau_C}\,(1-w^2)^{\frac{m+1}{2}}\right]^\frac{2}{3+m}\, dw,
\end{eqnarray}

\noindent where $w=v_{\rm obs}/v_\infty$ is the normalized velocity shift
across the line.  Figure~\ref{fig4b} shows curves for the EW normalized
to $\lambda_0 v_\infty/c$ and plotted as a function of the relative line
strength $\tau_L/\gamma(m)\tau_{\rm C}$.  The different curves are for different
values of $m$ at increments of 0.25.  Curves are closely clustered with
shallow slopes for larger values of $m$ representing steeper declines
in the clumping factor with radius in the wind.  

The overall result is that increasing values of $m$ steepen the decline
in clumping with radius, narrow the line width, increase the relative
line strength (because $\gamma$ decreases with $m$), and decrease the
line EW.  Equations~(\ref{eq:tS}) and (\ref{eq:tC}) show how the line
and continuum optical depths change with $m$, and the rate of decline
for these is not the same along a fixed sightline.  The line becomes
thinner more rapidly at a given resonance point as compared to the total
line-of-sight free-free optical depth.  

\section{Application to WR~90}	\label{sec:app}

With the solution to the radiative transfer through the wind at hand, it
is useful to consider an illustrative model.  The continuum slope is
set by equation~(\ref{eq:fluxscale}).  
To construct a full emission line
spectrum, a tabulation of line optical depth scales $\{ \tau_{L,{\rm i}}
\}$ are required as input to the model.  The IR spectra of WR stars are
characteristically He\,{\sc ii} emission line spectra, and at wavelengths
longer than about 10 microns (relevant to the {\it Spitzer}/IRS), the
line transitions involve somewhat high $n$-values, levels above $n=9$.

With the LTE assumption, the Sobolev optical depth scale factors
vary with the line transition as follows:

\begin{equation}
\tau_L \propto gf\,\lambda\, \left( \frac{n_i}{g_i} - \frac{n_j}{g_j}\right),
\end{equation}

\noindent where $f$ is the oscillator strength, $g$ values are statistical
weights for respective levels, and $n$ is a scaled number density (i.e.,
no dependence on radius).  Assuming the Rayleigh-Jeans limit as a simple
case, the final result of this is that

\begin{equation}
\tau_L \propto gf \, \frac{n_j}{g_j} \propto \lambda^2\,A_{ji}\, n_j.
\end{equation}

\noindent 
It is now easy to compare the line optical
scale factors for any two lines $\tau_L$ and $\tau_L'$ involving different
upper levels $j$ and $j'$. Given in ratio, the optical depth scales are
related as follows,

\begin{equation}
\frac{\tau_L'}{\tau_L} = \left(\frac{\lambda'}{\lambda}\right)^2\,
	\left(\frac{A'}{A}\right)\,
	\left(\frac{g'}{g}\right).
\end{equation}

\begin{figure}[t]
\centering{\epsfig{figure=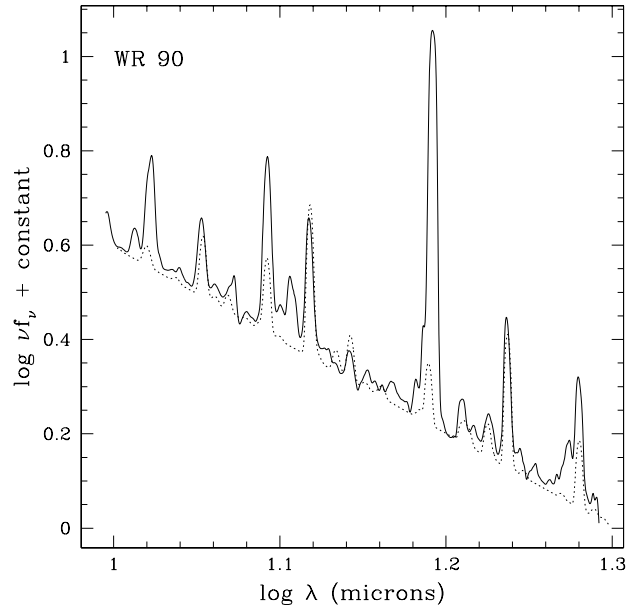, height=8.6cm}}
\caption{\small Shown is {\it Spitzer} data for the WC star, WR~90
(solid), along with a model fit using the FREC model
(dotted).  This is a two-parameter fit:  $m=0.9$ is
chosen to match the continuum slope, and the lines are governed by a
single optical depth parameter assuming LTE for relative level
populations.  The spectrum has been gaussian convolved to the resolution
of the IRS/SH instrument.  The stronger lines of He\,{\sc ii} are approximately
matched.  Note the prominence of [Ne\,{\sc iii}] 15.55 microns,
which is not part of the FREC and would need to be modeled separately
(e.g., Smith \& Houck 2005). \label{fig5}}
\end{figure}

A FREC spectral model depends on just three items: (1) $m$ that sets the
continuum slope, (2) $\tau_0$ a fiducial line optical depth for some
line from which all other lines scale, and (3) a list of He\,{\sc ii}
line data.  An example application in Fig.~\ref{fig5} shows a FREC model
overplotted with {\it Spitzer} IRS/SH data of the WC star, WR~90.  Data
are shown as solid, and the model is the dotted curve.  Transitions for
He\,{\sc ii} up to $n=30$ are included.  The observed spectrum includes
lines of other atomic species besides just recombination lines of He\,{\sc
ii}, notably the booming line of [Ne\,{\sc iii}] at 15.55 microns, but
also lines from ionized carbon.  The model makes no attempt to include any
lines other than He\,{\sc ii} recombination lines.  The dotted curve is
the simple two-parameter spectral model as ``eyeballed'' for a good fit.
The model does adopt the wind terminal speed of WR~90 from Ignace \etal\
(2007) and has been gaussian convolved to match the resolution of the
IRS in SH mode.

A value of $m=0.9$ reasonably reproduces the observed slope.  After
fixing the continuum slope, the optical depth scale was varied until
approximate matches to the observed emission in He\,{\sc ii} $10\alpha$
at 13.12~$\mu$m and $11\alpha$ at 17.26~$\mu$m were achieved.  The fit
is only illustrative, and no claim is made that $m=0.9$ is an accurate
description of the clumping distribution in the wind of WR~90.  Modeling
has shown that the winds of WC~stars have rather extended acceleration
zones to large radii (Hillier \& Miller 1999; Gr\"{a}fener \& Hamann
2005), hence the underlying assumption of constant radial expansion
in the region where the 10--20 micron emission forms may not be valid
for WR~90.  Still, this example application to {\it Spitzer} data shows
that the FREC model can give qualitatively reasonable fits.

\section{Summary}	\label{sec:summ}

In this paper an elegant result originally derived by Hillier \etal\
(1983) has been expanded significantly to allow for greater versatility.
Although the assumptions remain restrictive (spherical symmetry,
isothermal, and the constant radial expansion), the solution to
the radiative transfer is analytic and allows for both line blends
and a power-law dependence of the clumping with radius.  The latter
is especially significant since clumping is now understood to be an
important consideration in deriving mass-loss rates from observations
and since there are outstanding questions regarding the evolution of
the clumping factor throughout stellar winds.

What is especially interesting about the new FREC models is a relation
between line widths and line equivalent widths that has a dependence
on the exponent $m$.  Quantitatively the predicted relations may not
be realized in data owing to the various assumptions built into making
the FREC solution analytic; however, models based on the generalized
line source functions from equation~(\ref{eq:IL}) may be new avenues of
consideration for quick parameter exploration for those who use the more
sophisticated radiative transfer techniques for spectral synthesis of
stellar winds.  Archival data from the {\it Infrared Space Observatory}
or the {\it Spitzer} satellite may have adequate spectral resolution for
sources with fast winds to test against such relations and place limits
on the variation of clumping with radius.

Although this paper focused on He\,{\sc ii} emission line spectra of
WR~winds for application, owing to the high densities of these winds,
the solution may be applicable to other sources, like the hydrogen
recombination line spectra of Luminous Blue Variable (LBV) stars, such as
P~Cygni (Lamers \etal\ 1996).  The FREC solution may even have application
data to suitable hypercompact or ultracompact H\,{\sc ii} regions.  These
sources show radio recombination lines and free-free continuum spectra
plus evidence for density gradients (e.g., Keto, Qizhou, \& Kurtz 2008).

\section*{Acknowledgements}

I want to thank both the referee Wolf-Rainer Hamann and my good friend
Joe Cassinelli for a number of helpful comments.  I also gratefully
acknowledge Greg Tracy for his efforts in reducing the {\it Spitzer}
data of WR~90.

\end{document}